\documentclass[journal]{IEEEtran}

\usepackage{amsmath,amssymb}
\usepackage{multicol}
\usepackage{graphicx}
\usepackage{amsmath}
\usepackage{caption}
\usepackage{algorithm}
\usepackage{algpseudocode} 
\usepackage{placeins}
\usepackage{color}

\usepackage{xcolor}

\title{Signal and Backward Raman Pump Power Optimization in Multi-Band Systems Using Fast Power Profile Estimation  }
\author{Y. Jiang, J. Sarkis, S. Piciaccia, F. Forghieri,  
 P. Poggiolini
\thanks{Y. Jiang, J. Sarkis and P. Poggiolini are with Politecnico di Torino, Torino, Italy. S.~Piciaccia and F.~Forghieri, are with CISCO Photonics, Vimercate (MB), Italy.}

\thanks{This work was partially supported by: Cisco Systems through the BOOST research contract; the PhotoNext Center of Politecnico di Torino; the European Union under the Italian National Recovery and Resilience Plan (NRRP) of NextGenerationEU, partnership on “Telecommunications of the Future” (PE00000001 - program “RESTART”).}}
\date{February 2024}

\begin{document}

\maketitle
\begin{abstract}
This paper presents an efficient numerical method for calculating spatial power profiles of both signal and pump with significant Interchannel Stimulated Raman Scattering (ISRS) and backward Raman amplification in multiband systems. This method was evaluated in the optimization of a C+L+S/C+L+S+E 1000km link, employing three backward Raman pumps, by means of a closed-form EGN model (CFM6). The results show a 100x computational speed increase, enabling deep optimization which made it possible to obtain very good overall system performance and flat GSNR.
\end{abstract}
    

\section{Introduction}
\label{sec:Introduction}
Multiband systems have emerged as a solution to enhance the capacity of optical communication networks. The extension from the traditional C-band to C+L has been successfully implemented in commercial systems, effectively doubling system throughput. Ongoing research is focused on incorporating additional bands, such as S, U, and even E and O \cite{2024_PTL_Jiang}-\cite{2024_ECOC_Kim}. 

In multiband systems, Interchannel Stimulated Raman Scattering (ISRS) is very strong. Among the techniques that have been explored to mitigate the impact of ISRS, launch power spectrum optimization and selective Backward Raman amplification have been shown to be quite effective \cite{2024_PTL_Jiang}. However, such techniques are typically based on iterative algorithms and, to be practical, they need to be supported by very fast physical layer models. The traditional numerically-integrated GN and EGN models are too slow, by several orders of magnitude. Closed-form models (CFMs) are needed.

Two research groups, one at University College London (UCL), and one at Politecnico di Torino (PoliTo) in collaboration with CISCO, have independently developed over the last few years fast NLI (Non-Linear Interference) CFMs, based on the GN and EGN physical layer models. For the UCL CFM see \cite{2023_JLT_Buglia}, \cite{2024_JLT_Buglia}, for the CISCO-PoliTo CFM see \cite{2022_ECOC_Poggiolini}, \cite{2024_ECOC_Jiang}. 

These CFMs are highly effective, to the extent that they are no longer the primary computational overhead. Instead, the dominant factor has become Spatial Power Profile (SPP) estimation. The CFMs need the SPP of each channel along each span to produce accurate NLI estimation. 

The computation of the SPPs is typically based on numerically solving the coupled differential Raman equations, which account for both ISRS and Raman amplification \cite{2002_PTL_Perlin}. Conventional numerical methods, such as MATLAB ODE solvers, handle such computation efficiently when ISRS alone is present. However, when \textit{backward-pumped} Raman amplification is introduced, the problem transforms into a dual-boundary-condition one, which is much slower to resolve. SPP estimation then becomes the dominant computational bottleneck, consuming 95\% or more of the total time required to evaluate overall system throughput.

Several approaches have been proposed to expedite SPP calculations. Some rely on approximations, such as assuming a triangular-shaped Raman gain spectrum. While fast, this method is affected by substantial error. Other approaches try to speed up computation by modifying the integration algorithm. For instance, \cite{2024_OFC_Kimura} finds a coarse SPP solution by assuming frequency-independent pump depletion, and then takes such solution as a starting point to find the actual one. However, the speed-up achieved with this method is relatively modest.

Another method \cite{2005_OL_Choi} focuses exclusively on \textit{backward} propagation, projecting the forward direction onto the backward. It employs vectorization techniques to enhance computational efficiency. However, while this method is very fast when it converges, it has frequent convergence failures which lead to unreliable performance, limiting its practical applicability. 

Nonetheless, we found \cite{2005_OL_Choi} very promising and in this paper we developed a similar algorithm that both further improved computational efficiency and solved the convergence issues. Similar to \cite{2005_OL_Choi}, propagation is unidirectional. Differently from it, we focus exclusively on \textit{forward} propagation. We also devised suitable techniques to prevent divergence. Eventually, we got approximately a 200x speed increase vs. conventional ODE algorithms, while consistently ensuring convergence. 

As test case-studies for the algorithm, we looked at long-haul SMF systems operating in the C+L+S (CLS for short) and C+L+S+E (CLSE for short) bands, with backward Raman amplification. Both the WDM launch power spectrum and backward Raman amplification are optimized, first to maximize throughput and then also to ensure GSNR flatness. Thanks to the deep optimization made possible by the CFM and the sped-up SPP estimation algorithm, we obtain excellent overall system performance, combining high throughput with flat GSNR. 

This paper is a follow-up to \cite{2025_OFC_Sarkis}. With respect to \cite{2025_OFC_Sarkis} we have revised and improved the algorithm, greatly expanded on its description and completely recalculated all results after adding double Rayleigh back-scattering, which is a non-negligible contribution to noise when backward Raman amplification is significant. We have explicitly reported the objective functions used for optimization and discussed the different weights that can be used for achieving different levels of GSNR flatness. CLS results are new, as well as CLSE with GSNR flatness. Only CLSE for maximum throughput was already in \cite{2025_OFC_Sarkis}, but without DBR. We have also added full EGN-model checks of the final CFM-based results.

This paper is organized as follows. In Sect.~\ref{sec:SPPs_algorithm} the fast power profile algorithm is introduced. In Sect.~\ref{sec:System_and_Results} the system is described in detail and GSNR results are reported and extensively discussed. Comments and conclusion follow.

\section{The Unidirectional Algorithm}
\label{sec:SPPs_algorithm}
The SPP along a span of a channel or a backward Raman pump (BRP) can be found using the well-known coupled Raman differential equations. In the following, please keep in mind that we call ‘lightwaves’ both the individual transmission channels and the individual BRPs. We identify each lightwave through an index $n$ associated to its center frequency $f_n$. Note that both transmission channels and BRPs can actually act as ‘pumps’, that is they can feed power to another lightwave, or as ‘probes’, that is they can receive power from another lightwave.

Systems can also make use of forward Raman Pumps (FRPs), though this is a niche technique used on rare occasions. In the following we will not mention them explicitly. If present, FRPs are treated by the algorithm exactly as the forward-propagating transmission channels.

For the $n$-th lightwave, its power profile  ${P_n}\left( z \right)$ is given by:
\begin{equation}
    \begin{aligned}
       \pm \frac{dP_n(z)}{dz} & = 
        - 2 \cdot \alpha_{n} \cdot P_n(z) \\
       & + \left\{ \sum_{j=1}^N \varsigma \left( f_n,f_j \right) \cdot C_R(f_n, f_j) \cdot P_j(z) \right\} \cdot P_n(z) 
    \end{aligned}
    \label{eq:diff_power_profile}
\end{equation}
where the upper sign in the left is used when $P_n (z)$ is a forward-propagating lightwave, the lower sign is used when it is a counter-propagating lightwave. Note that in practical systems the counter-propagating lightwaves consist of just BRPs, whereas the forward-propagating lightwaves consist of channels and possibly FRPs as well. Such equations can be re-written in integral form:         
\begin{equation}
    \begin{aligned}
        P_n(z) = & P_n(0) \cdot \exp \Bigg\{ 
        \mp 2 \cdot \alpha_{n} \cdot z \\
        & \pm \sum_{j=1}^N \varsigma \left(f_n,f_j \right) \cdot C_R(f_n, f_j) \cdot \int_0^z P_j(\xi) \, d\xi \Bigg\}
    \end{aligned}
    \label{eq:int_power_profile}
\end{equation}
where $N$ is the total number of lightwaves, $\alpha_{n}$  represents the intrinsic fiber loss at the frequency of the $n$-th lightwave and $C_R (f_n,f_j )$ represents the gain/loss coefficient experienced by the $n$-th lightwave due to the presence of the $j$-th lightwave. The function $\varsigma \left( {f_n},{f_j} \right)$ is $\left({f_n} / {f_j}\right)$ when ${f_n} > {f_j}$, 1 when ${f_n} < {f_j}$ and 0 when ${f_n} = {f_j}$.  

\begin{figure*}
    \centering
    \includegraphics[width=0.75\linewidth]{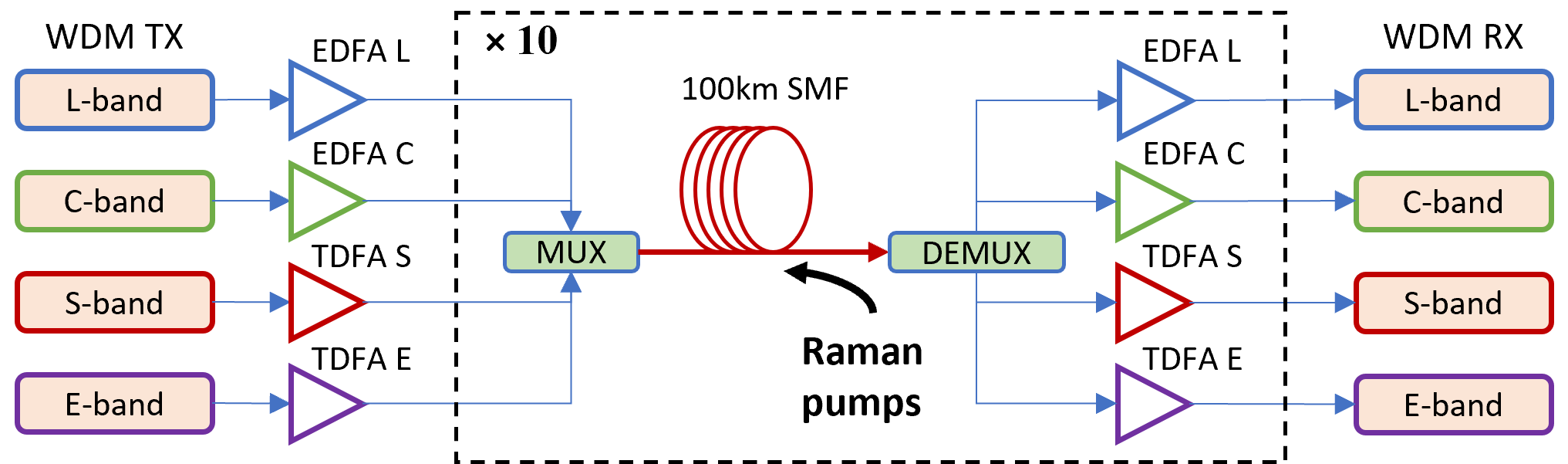}
    \vspace{0.5cm}
    \caption{Schematic of the 10-span CLSE system being studied. Fiber parameters are slightly different among spans (see text).}
    \label{fig:schematic}
\end{figure*}

The system of $N$ coupled equations Eq.~(\ref{eq:int_power_profile}) is a double-boundary value problem, which typically needs many forward-backward iterations to achieve a solution starting from a reasonable guess. One alternative is \cite{2005_OL_Choi} which, though iterative as well, only resorts to \textit{backward} solutions, that can be put into a very computationally efficient matrix form, dealing with all the coupled equations at once. We tried this approach but, as mentioned in the Sect.~\ref{sec:Introduction}, we found substantial convergence problems. 

We decided therefore to focus instead on \textit{forward}-only propagation, to see if we could improve reliability. We also used a matrix form to speed up computations. Calling $\mathbf{P}$ the matrix whose rows are the spatial samples of the $n$-th lightwave SPP, taken at $z=m \cdot \Delta z$ along the span in the forward direction,  that is, \( \mathbf{P} = [P_n(m \cdot \Delta z)]_{N \times M} \), we can generate the \((k+1)\)-th power-profile iteration from the \(k\)-th as follows:
\begin{equation}
    \mathbf{P}^{(k+1)} = \mathbf{P}_{\rm in}^{(k)} \cdot \exp \left\{ 
        \mp 2 \cdot \boldsymbol{\alpha} \cdot \mathbf{z} 
        \pm \Delta z \cdot \mathbf{C}_R \cdot \mathbf{P}^{(k)} \cdot \mathbf{U} 
    \right\}
    \label{eq:matrix_power_profile}
\end{equation}
where:
\begin{equation}
    \begin{aligned}
     & \mathbf{P}_{\rm in}^{(k)} = \text{diag} \left( \left( P_n^{(k)}(0) \right)_{N \times N} \right) \\[0.2cm]
     & \boldsymbol{\alpha}  = \left( \alpha_{n} \right)_{N \times 1}\\[0.2cm]
     & \mathbf{z} = \left( m \cdot \Delta z \right)_{1 \times M} \;\; , \;\; \Delta z = \frac{L_{\text{span}}}{M} \\[0.2cm]
    & \mathbf{C}_R = \left[ \varsigma \left( f_n,f_j \right) \cdot C_R(f_n,f_j) \right]_{N \times N}\\[0.2cm]
    & \mathbf{U} = \left[ \begin{array}{*{20}{c}}
    0 & \frac{1}{2} & \frac{1}{2} & \frac{1}{2} & \dots & \frac{1}{2} \\
    0 & \frac{1}{2} & 1 & 1 & \dots & 1 \\
    0 & 0 & \frac{1}{2} & 1 & \dots & 1 \\
    0 & 0 & 0 & \frac{1}{2} & \dots & 1 \\
    \vdots & \vdots & \vdots & \vdots & \ddots & \vdots \\
    0 & 0 & 0 & 0 & \dots & \frac{1}{2}\end{array} \right]_{M \times M} \\[0.2cm]
    \end{aligned}
    \label{eq:uni_power_profile1}
\end{equation}
where $L_{\text{span}}$ is the span length. Note that the product $\left( \Delta z \cdot \mathbf{P} \cdot \mathbf{U} \right)$ in Eq.~(\ref{eq:matrix_power_profile}) performs numerical integration according to the trapezoidal method, in the forward direction. Note also that even the BRPs are fictitiously propagated forward by Eq.~(\ref{eq:matrix_power_profile}). This is mathematically possible by flipping the sign of the intrinsic loss $\alpha_n$ and Raman gain $C_R$ in Eqs.~(\ref{eq:int_power_profile}) and (\ref{eq:matrix_power_profile}) for all lightwaves that physically back-propagate, but are dealt with by the algorithm as forward-propagating.

At the first iteration, Eq.~(\ref{eq:matrix_power_profile}) needs to be initialized with a guess of the SPPs for all lightwaves, across the entire span: \( \mathbf{P}^{(0)} \). This matrix also directly implies the boundary condition at ($z\!=\!0$), that is $\mathbf{P}_{\rm in}^{(0)}$. The initialization of \( \mathbf{P}^{(0)} \) is a critical aspect of the algorithm and will be discussed in Sect.~(\ref{sect:scaling}). 


Once \( \mathbf{P}^{(0)} \) has been initialized, then Eq.~(\ref{eq:matrix_power_profile}) generates the next iteration, \( \mathbf{P}^{(1)} \), which in turn generates \( \mathbf{P}^{(2{
})} \), and so on. Ideally, the algorithm has converged when \textit{both} \( \mathbf{P}^{(k+1)} = \mathbf{P}^{(k)} \) \textit{and} the power of each BRP at $z\!=\! L_{\rm span}$ coincides with its actual nominal launch power value.

 After substantial testing, we found that this algorithm had better reliability, but still convergence failed occasionally. Also, there were limitations as to the pump power that it was capable to handle. Therefore, we improved the algorithm by devising suitable strategies to both assign the initial condition \( \mathbf{P}^{(0)} \) and to adapt the SPPs fed to Eq.~(\ref{eq:matrix_power_profile}) at \textit{each successive iteration}. The goal was to ensure that each iteration generated a solution \( \mathbf{P}^{(k)} \) closer to the actual accurate result, while avoiding divergence, oscillations, or excessive iterations. 

\subsection{Modifying the algorithm}
\label{sect:scaling}

The initial SPPs are imposed as follows. The power profiles of all lightwaves are calculated based only on fiber loss. Then, the SPPs of the BRPs are rescaled, according to this simple criterion. If the sum of the launch powers of the BRPs at $z\!=\!L_{\rm span}$ is larger than the sum of the launch powers of the transmission channels at $z\!=\!0$, then the BRPs SPPs are scaled down so that the two sums coincide. Otherwise, they are left unchanged.

However, the goal of the algorithm is to find a solution where the boundary conditions on launch powers are satisfied both at $z\!=\!0$ for the transmission channels and at $z\!=\!L_{\rm span}$ for the BRPs. We pre-define a series of power references ${P}_\text{pump,ref}^{(k)}$ at $z\!=\!L_{\rm span}$ to gradually scale BRPs from initial powers up to the launch powers by small steps. Smaller steps are taken when the difference between nominal BRP launch powers and the ones after the $k$-th iteration is smaller. More details are reported in the Appendix. Then the algorithm starts.

All initial SPPs are loaded into \( \mathbf{P}^{(0)} \) and a first integration solution \( \mathbf{\tilde P}^{(1)} \) is produced. The solution \( \mathbf{\tilde P}^{(1)} \) in general contains values of the BRPs at $z\!=\!L_{\rm span}$ that are different from ${P}_\text{pump,ref}^{(1)}$. This happens even if the initial BRP rescaling described above was not performed, where ${P}_\text{pump,ref}^{(1)}$ is the BRPs nominal launch powers. The reason is the forward-nature of the algorithm, which does not allow to impose any constraints on the SPP values at $z\!=\!L_{\rm span}$.

Before re-inserting \( \mathbf{\tilde P}^{(1)} \) into Eq.~(\ref{eq:matrix_power_profile}), BRPs SPPs are modified to match ${P}_\text{pump,ref}^{(1)}$ at $z\!=\!L_{\rm span}$. This creates \( \mathbf{ P}^{(1)} \), that is passed into Eq.~(\ref{eq:matrix_power_profile}), producing \( \mathbf{\tilde P}^{(2)} \). Again, the BRP SPPs are scaled up/down slightly according to the same law and a new matrix  \( \mathbf{ P}^{(2)} \) is obtained and passed through Eq.~(\ref{eq:matrix_power_profile}) to obtain \( \mathbf{\tilde P}^{(3)} \), and so on. 

As the iterations keep going, the gap between the BRP launch power in the fiber at the algorithm $k$-th step \( \mathbf{\tilde P}^{(k)} \) is gradually reduced versus the {\it actual} BRP launch power. The algorithm stops when all BRPs launch powers are back to their nominal values and no significant fluctuations are detected in the transmission channels SPPs.

Using this feedback-guided scaling strategy, we have been able to obtain very reliable convergence for the algorithm, which we have tested in very high pump and channel power situations, well beyond practical levels, with excellent results. Despite the algorithm requiring typically 50 to 200 iterations, due to its efficient matrix formulation and inherent features, it has consistently proved about 200x faster than our Matlab reference algorithm in SPPs calculation only. We will come back on performance testing in Sect.\ref{sect:performance}. A detailed description and testing of the algorithm can be found in \cite{2024_arXiv_Sarkis}.

\section{System Case Studies}
\label{sec:System_and_Results}

We tested the sped-up algorithm on the optimization of a CLS and a CLSE 10-span system, with BRPs in the higher-frequency bands. The use of BRPs has proven quite beneficial for such multiband links.

These systems require complex launch power optimization, channel by channel, as well as Raman pump (power and frequency) optimization. In turn, this entails hundreds or thousands of full-system evaluations, which would be very difficult to carry out without a CFM and, due to the presence of BRPs, without an efficient SPP calculation algorithm. They are therefore ideal for testing the proposed SPP calculation algorithm.

Incidentally, while using this as a case-study for testing the SPP algorithm, we obtain interesting system results that have their own validity and impact. We will comment on the resulting system indications at the bottom of the paper.

\begin{figure}
    \centering
    \includegraphics[width=0.9\linewidth]{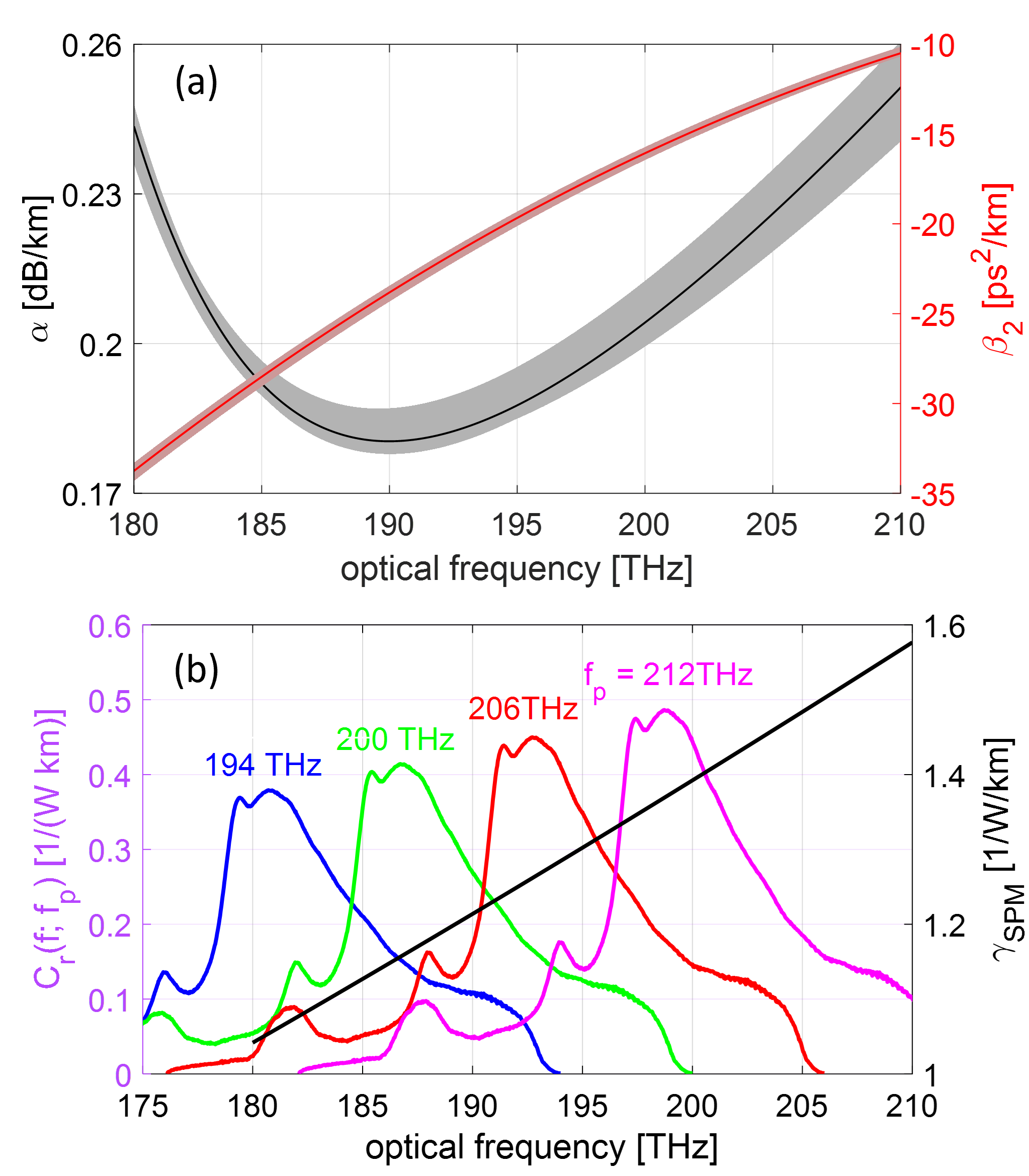}
    \caption{(a) Loss and dispersion, measured in C and L bands and extrapolated to S and E bands. (b) Raman Gain spectrum and SPM coefficient $\gamma$.}
    \label{fig:loss_and_dispersion_Raman_gamma}
\end{figure}
\begin{figure}
    \centering
    \includegraphics[width=0.95\linewidth]{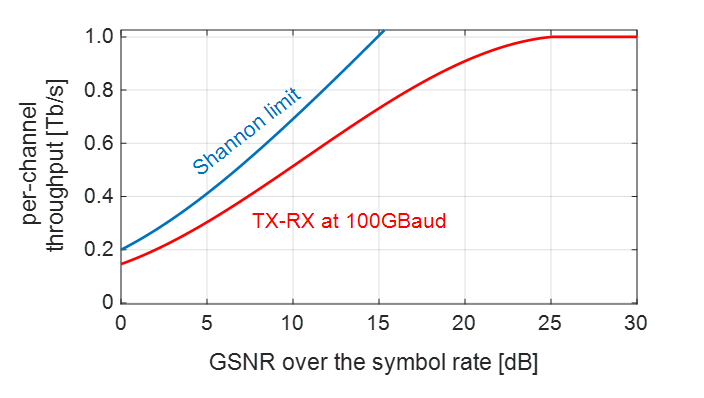}
    \caption{Transceiver net per-channel throughput vs.~GSNR.}
    \label{fig:IR}
\end{figure}

\subsection{System description}
The schematic is shown in Fig.~\ref{fig:schematic}. The first five spans, all slightly different from one another, were characterized from the experimental set-up used for CFM validation in \cite{2024_JLT_Jiang}. Loss and dispersion were measured in C and L band and then extrapolated to the S and E band (Fig.~\ref{fig:loss_and_dispersion_Raman_gamma}(a)) using well-known formulas \cite{1986_JLT_Walker}. The Raman gain spectrum $C_R(f, f_p)$ was experimentally characterized using a pump at $f_p$=206.5~THz (Fig.~\ref{fig:loss_and_dispersion_Raman_gamma}(b)). It was shifted and scaled as a function of $f$ and $f_p$, according to \cite{2003_JLT_Rottwitt}. The black line in Fig.~\ref{fig:loss_and_dispersion_Raman_gamma}(b) shows the non-linearity coefficient $\gamma$ vs.~frequency. 

The first five spans were then extrapolated to 100~km each and replicated, to achieve a total system length of 1000~km. The insertion loss from the MUX/DEMUX and connectors was assumed to be 4~dB per span. The average total span loss was 22dB at 193THz. 

The WDM band boundaries set here were: L-band 184.50 to 190.35; C-band 190.75 to 196.60; S-band 197.00 to 202.85; E-band 203.25 to 209.07~THz. Doped-Fiber-Amplifiers (DFAs) were assumed, with a noise figure of 5~dB in the C-band, 6~dB in the L and S-bands, and 7~dB in the E-band. Each band contained 50 equally-spaced channels, with a symbol rate of 100~GBaud, roll-off 0.1, and spacing of 118.75~GHz. The modulation was taken to be Gaussian-shaped. The \textit{net} per-channel throughput of the transceivers, vs.~GSNR at the receiver input, was assumed as shown in Fig.~\ref{fig:IR}. BRPs were included, consisting of three pumps.

The system was then optimized. Specifically, the launch power spectrum in each band and the frequency and power of the three BRPs were optimized first for \textit{maximum throughput}, and then GSNR (Generalized Signal-to-Noise Ratio, Eq.~(\ref{eq:GSNR_DRB})) flatness. Note that the launch power spectrum was described using a cubic polynomial in each band: 
\begin{equation}
    P_{\rm{ch,dB}}(f) = a_0 + a_1(f - f_c) + a_2(f - f_c)^2 
    + a_3(f - f_c)^3
    \label{eq:power_spectrum}
\end{equation}
with \( f_c \) being the center frequency of each band. This spectrum representation requires a total of 12 coefficients for CLS and 16 coefficients for CLSE systems (4 per band) and significantly simplifies the optimization problem as compared to a per-channel power optimization. The launch power spectrum is assumed to be the same at the start of each span, assuming the presence of programmable optical filters in DFAs, as it is commonplace in latest-generation units. 

The system GSNR is defined as:
\begin{equation}
    \text{GSNR} = \frac{P_{\text{ch}}}{P_{\text{ASE}} + P_{\text{NLI}} + P_{\text{DRB}}}
    \label{eq:GSNR_DRB}
\end{equation}
where $P_{\text{ch}}$ is the power of the channel being tested and $P_{\text{ASE}}$, $P_{\text{NLI}}$,  $P_{\text{DRB}}$ are the power of ASE noise, of NLI noise, and of double Rayleigh Back-Scattering (DRB) noise \cite{2002_ECOC_Dibon}, respectively.
$P_{\text{DRB}}$ is obtained by summing up the $P_{\text{DRB},n_\text{s}}$ from each span. $P_{\text{DRB},n_\text{s}}$ is expressed as,
\begin{equation}    
    \begin{aligned}
        P_{\text{DRB},n_\text{s}} & = G_{\text{DFA},n_\text{s}} \times \\
        & \sum_{z_1=2}^M \sum_{z_2=1}^{z_1-1} \left( P_{\mathrm{ch}} \times G_{1, M} \times \Delta z^2 \times \kappa^2 \times   G_{z_1 z_2}^2 \right)    
    \end{aligned}
    \end{equation}
where $G_{\text{DFA},n_\text{s}}$ is the DFAs gain at the end of $n_\text{s}$-th span to ensure the same launch power across all spans . $\kappa$ is Rayleigh back-scattering coefficient, assumed to be -40~dB/km across all channels, $G_{z_1 z_2}$ is the gain/loss from $z_1$ to $z_2$. In this paper, it is assumed that the span loss of each channel is fully and exactly compensated for at the end of the span, taking ISRS into account. Otherwise, $P_{\text{DRB},n_\text{s}}$ should be propagated linearly along the subsequent spans towards the receiver. 

\begin{figure}
    \centering
    \includegraphics[width=1.0\linewidth]{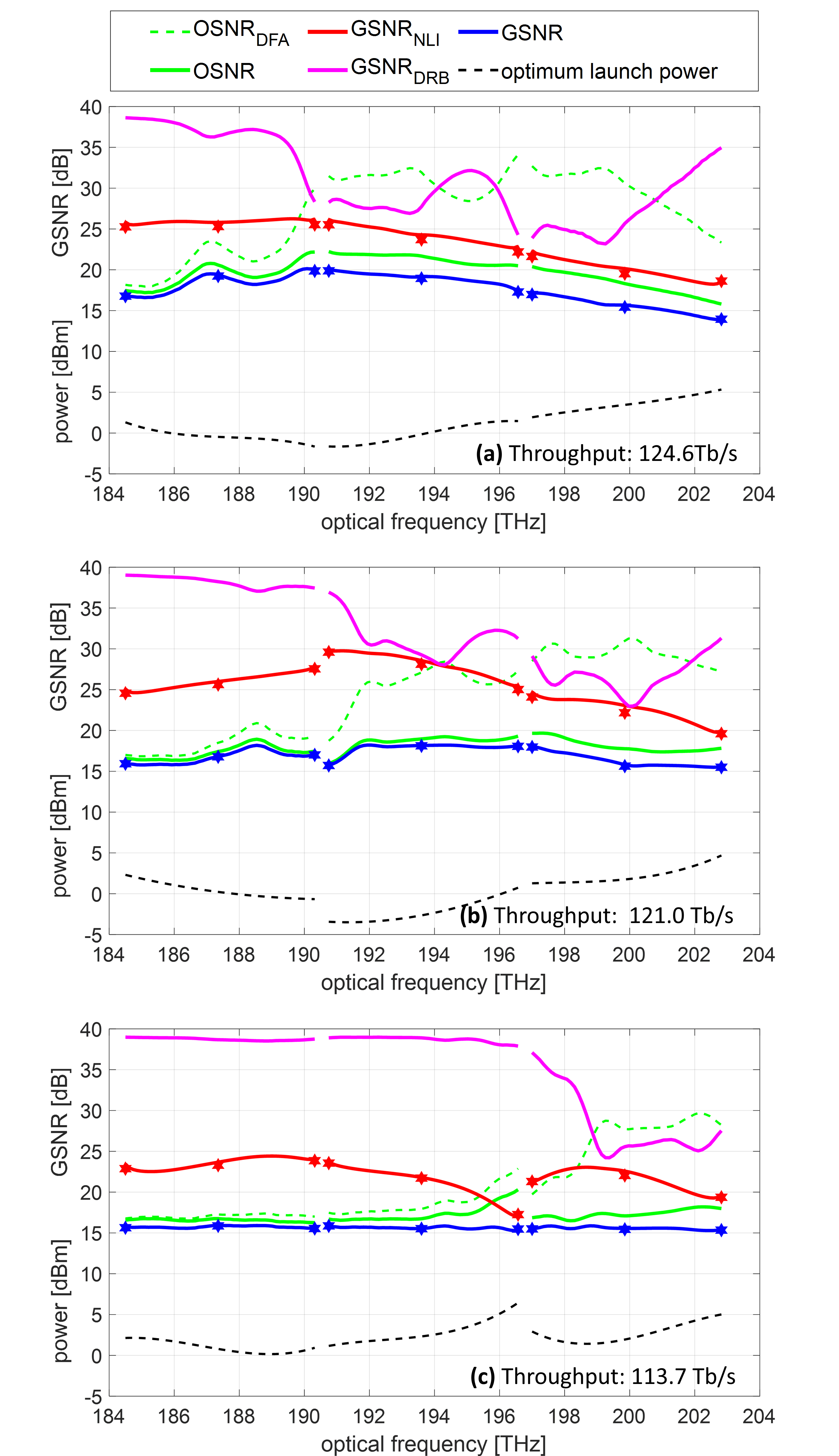}
    \caption{CLS optimum launch power per channel and GSNRs for (a) throughput maximization using Eq.~(\ref{eq:objective_function_maxThroughput}), the optimized BRPs settled at: 205.1 THz, 21.5 dBm; 211.5 THz, 27.7 dBm; 214.0 THz, 26.6 dBm; (b) throughput maximization while flattening GSNR using Eq.~(\ref{eq:objective_function_flat}) with $w$ = 0.5, the optimized BRPs settled at: 206.7 THz, 21.3 dBm; 212.4 THz, 26.6 dBm; 214.8 THz, 26.8 dBm. (c) throughput maximization while flattening GSNR using Eq.~(\ref{eq:objective_function_flat}) with $w$ = 1, the optimized BRPs settled at: 212.4 THz, 22.2 dBm; 214.0 THz, 25.2 dBm; 217.0 THz, 25.5 dBm. Here, all BRPs are constrained to be at least 2THz away from the signals.} 
    \label{fig:3bands_max_throughput}
    
\end{figure}

\begin{figure}
    \centering
    \includegraphics[width=0.9\linewidth]{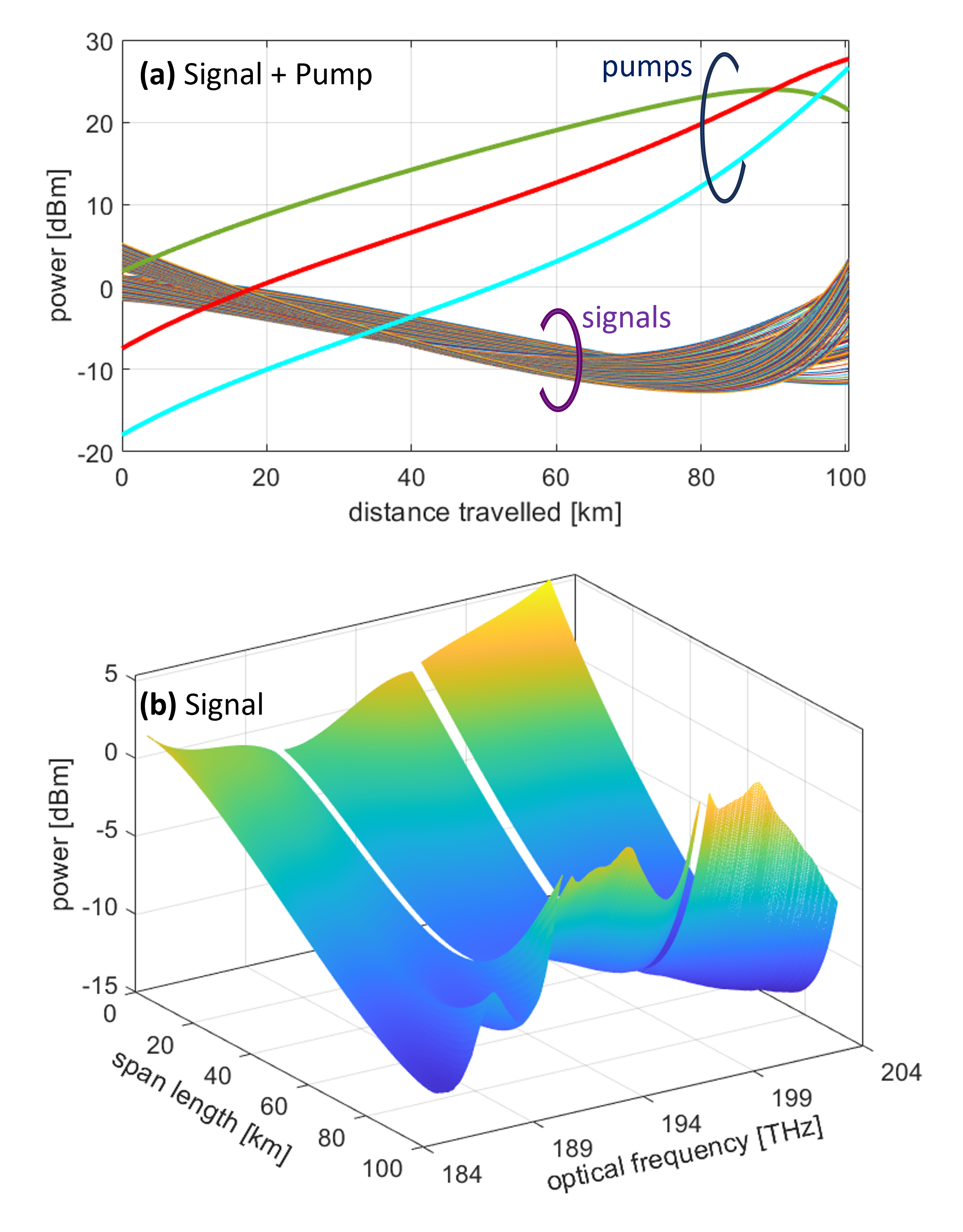}
    \caption{SPPs in the first span (a) 2D plot of all lightwaves; (b) 3D plot of all signals in the CLS system of Fig.~\ref{fig:3bands_max_throughput} (a).}
\label{fig:3bands_SPP}
    \vspace{-0.1cm}
\end{figure}

\begin{figure}
    \centering   
    \includegraphics[width=1.0\linewidth]{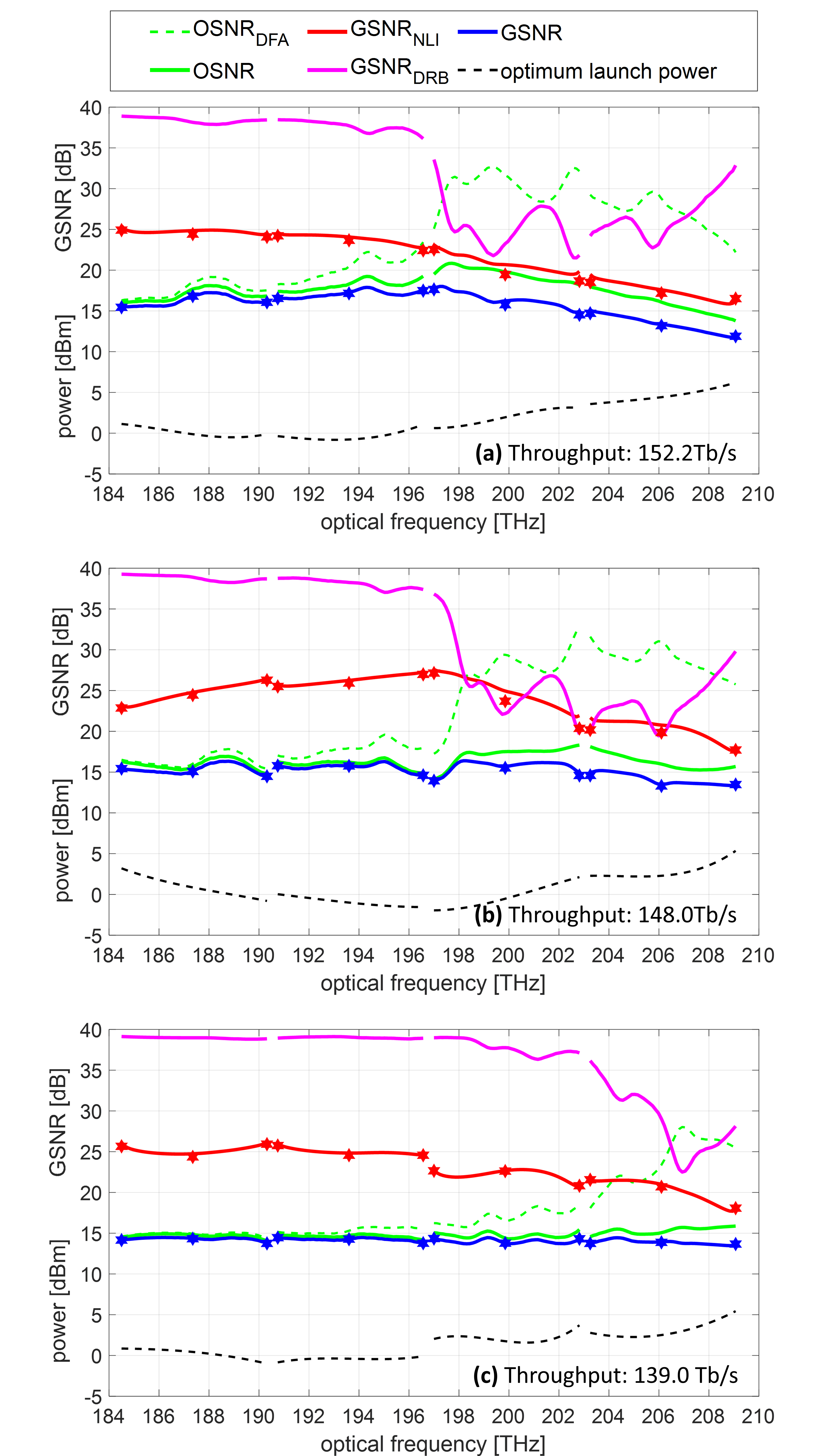}
    \vspace{0.0cm}
    \caption{CLSE optimum launch power per channel and GSNRs for (a) throughput maximization using Eq.~(\ref{eq:objective_function_maxThroughput}), the optimized BRPs settled at: 212.4 THz, 22.6 dBm; 217.5 THz, 25.7 dBm; 220.6 THz, 28.7 dBm; (b) throughput maximization while flattening GSNR using Eq.~(\ref{eq:objective_function_flat}) with $w$ = 0.5, the optimized BRPs settled at: 213.1 THz, 22.2 dBm; 217.8 THz, 25.7 dBm; 220.8 THz, 28.9 dBm; (c) throughput maximization while flattening GSNR using Eq.~(\ref{eq:objective_function_flat}) with $w$ = 1, the optimized BRPs settled at: 214.0 THz, 17.4 dBm; 219.2 THz, 24.0 dBm; 221.7 THz, 27.9 dBm. Here, all BRPs are constrained to be at least 2THz away from the signals.}
    \label{fig:4bands_max_throughput}
\end{figure}

\begin{figure}
    \centering
    \includegraphics[width=0.9\linewidth]{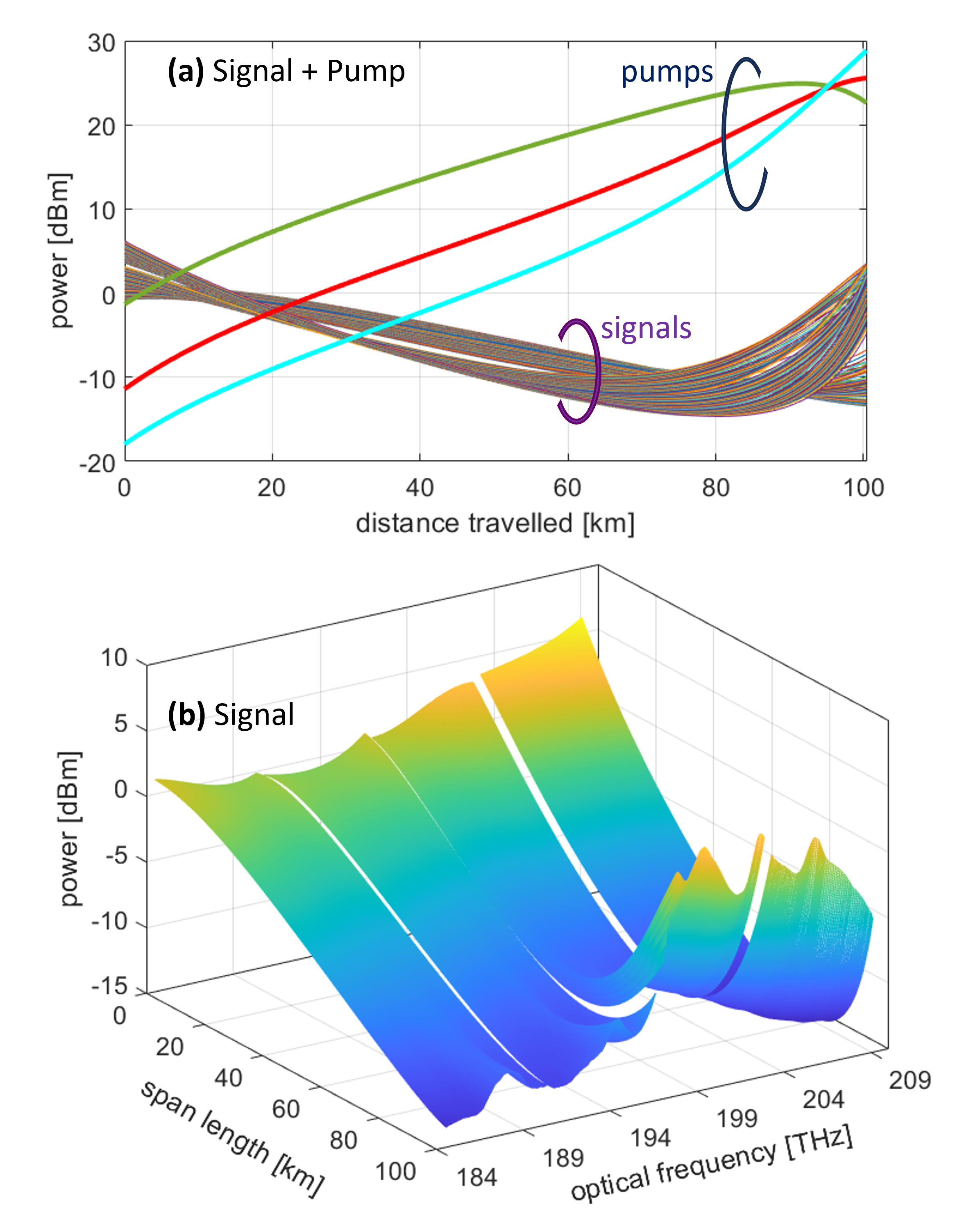}
    \caption{SPPs in the first span (a) 2D plot of all lightwaves; (b) 3D plot of all signals in the CLSE system of Fig.~\ref{fig:4bands_max_throughput} (a).}
\label{fig:4bands_SPPs}
\end{figure}

In Fig.~\ref{fig:3bands_max_throughput} (a) we show the CLS system results, with throughput maximization. The objective function is:
\begin{equation}
    f_{\rm obj} = {\rm mean}\left( {\rm T}_{\rm Rx}^{n} \right)   
    \label{eq:objective_function_maxThroughput}    
\end{equation}   
where ${\rm T}_{\rm Rx}^{n}$ is the net throughput of the $n$-th channel, obtained from the GSNR through the red curve in Fig.~\ref{fig:IR}. While not specific to any part number of any vendor, Fig.~\ref{fig:IR} is loosely representative of the performance expected of top-of-the-line transceivers using probabilistic shaping, operating at about 100~GBaud.

Fig.~\ref{fig:3bands_max_throughput} displays total $\text{GSNR}$ as well as various partial signal-to-noise ratios: $\text{OSNR}$ (ASE only, DFAs+Raman), $\text{GSNR}_\text{NLI}$ (NLI only), $\text{GSNR}_\text{DRB}$ (DRB only). $\text{OSNR}_\text{DFA}$ accounts for ASE from DFAs but not from BRPs. The much higher value of $\text{OSNR}_\text{DFA}$ in the C and S bands, as compared to $\text{OSNR}$, means that most amplification was delivered by BRPs in those bands. This can also be seen from the SPP plots in Fig.~\ref{fig:3bands_SPP}. The channels in both C band S bands experienced high gain from BRPs. This introduced more DRB noise in these two bands. Also, notice that all C and S channels are technically in nonlinear regime, that is, with
${\rm GSNR_{NLI} - OSNR} < 3 {\rm dB}$ \cite{2011_OFC_Bosco}.

Next, GSNR flatness was incorporated into the objective function to reduce GSNR variations, while still maximizing throughput. The objective function is modified as:
\begin{equation}
    f_{\rm obj} = {\rm mean}\left( {\rm T}_{\rm Rx}^{n} \right)-w *\mid {\rm T}_{\rm Rx}^{\max}-{\rm T}_{\rm Rx}^{\min} \mid    
    \label{eq:objective_function_flat}    
\end{equation} 
where $w$ is a weight used for controlling the flatness of GSNR. A bigger $w$ induces a flatter GSNR. 

Fig. ~\ref{fig:3bands_max_throughput} (b) shows the results with $w$ = 0.5.  The peak-to-peak GSNR variation was reduced to  2.7 dB from 6.2 dB, while the throughput went marginally down to 121.0 Tb/s from 124.6 Tb/s in Fig.~\ref{fig:3bands_max_throughput} (a). All BRPs (frequency and power were shown in the caption of Fig.~\ref{fig:3bands_max_throughput}) were moved about 1 THz farther away from the signal, by the optimization, resulting in an improvement of the performance in the higher-S band, but they still provided some amplification to C band. The GSNR in the highest-frequency channel goes up to 15.5 dB, gaining 1.6 dB compared to Fig.~\ref{fig:3bands_max_throughput} (a).

To achieve an even flatter GSNR,  $w$ was set to 1 in Fig.~\ref{fig:3bands_max_throughput} (c). The peak-to-peak GSNR variation was only 0.7 dB but the throughput experienced a bigger reduction, down to 113.7 Tb/s, about 10\% lower than the value found when only maximizing throughput. Notice that now propagation is mostly linear with only about 1/3 of the C-band channels and 1/4 of the S-band channels in non-linear regime. All BRPs were moved farther away from the signal, by the optimization, to concentrate on S-band amplification, making the DRB noise negligible in both C and L bands.   

We then looked at a CLSE system. Maximum throughput optimization achieved 152.2 Tb/s, a 22.2\% growth from CLS. The GSNR results are shown in Fig.~\ref{fig:4bands_max_throughput}(a) and SPPs plots are shown in Fig.~\ref{fig:4bands_SPPs}. Note that about 1/2 of the channels are in non-linear regime, where DRB noise is also significant. We then again introduced GSNR flatness in the objective function, reducing GSNR variation from 6.2 dB to 3.0 dB  with $w$ = 0.5 in (b) and 0.8 dB  with $w$ = 1 in (c), at 148.0 Tb/s (b) and 139.0 Tb/s (c) throughput in Fig.~\ref{fig:4bands_max_throughput}. As BRPs went further away from the signal, more and more channels were in linearity, and DRB noise became negligible. 

Note that other optimization solutions can be found by changing the weight of GSNR flatness in the objective function, resulting in a different trade-offs between total throughput and GSNR flatness. 

These case-studies were conceived to provide a demonstration of the effectiveness of the SPP algorithm in a very complex multiband long-haul environment. We think that overall they do provide interesting results related to multiband system design and optimization, but the key aspect is algorithm performance, which is dealt with below.

\subsection{Performance of algorithms}
\label{sect:performance}
Fig.~\ref{fig:4bands_SPPs} provides a picture of the very diverse SPPs of the different channels in the CLSE system of Fig.~\ref{fig:4bands_max_throughput}(a), which constituted a very challenging scenarios for the SPP estimator, as well as the CFM6 and the optimizer algorithm itself.

To double check the accuracy of the CFM6 results, for some of the channels we ran the numerically-integrated full-fledged EGN model on the final optimized system configurations, obtaining the star markers shown in the figures. Comparing markers with solid lines, CFM6 shows very good accuracy, even in difficult high non-linearity situations, with heavy ISRS and strong backward Raman amplification. As expected, CFM6 turned out to be absolutely essential, since the optimizations shown above needed on average 2000 full system evalutions each, which would be very challenging to run with numerically integrated NLI models.

CFMs solve the NLI estimation problem but system optimization would not have been possible without the key boost provided by the SPP evaluation algorithm. It improved SPP computation speed by a factor of about 200 when compared to the reference dual-direction Raman solver (bvp4c in MATLAB). As for its accuracy, we looked at 1,030,000 channel and pump power values, sampled along the links at 100m intervals, as found by the bvp4c algorithm and by our new SPP method. We found a max difference of ±0.02dB, proving the very good accuracy of the algorithm.

Regarding speed, one complete calculation of GSNR across 200 channels with CFM6, for the system of Fig.~\ref{fig:4bands_max_throughput}(a), took about 12 seconds. It is implemented in MATLAB(TM) and executed on a desktop PC equipped with a 12th Gen Intel(R) Core(TM) i9-12900 processor, operating at a clock speed of 2.40GHz. We did not make use of any GPU support in the calculations. It was distributed as follows: SPP assessment and NLI evaluation with the CFM, 21\% each; DRB estimation, 58.0\%. Notably, DRB estimation appears to be the most time-consuming component. However, we made no attempt to speed up DRB estimation. We believe it has large margins of improvement and is not the focus of this paper. 

In comparison, when using the conventional methods for SPPs calculation, one complete system performance calculation took 475 seconds, with 98.0\% for SPPs, 0.5\% for CFM and 1.5\% for DRB assessment. This shows that the fast SPP algorithm grants a 40x speed improvement overall. If DRB noise was not taken into account, or was made considerably faster, the overall speed improvement would 100x. Regarding SPP evaluation alone, the speed up is a remarkable 200x.

\section{Conclusion}
Fast full system performance estimation is essential for the deep iterative optimization of multiband systems. Its two time-consuming steps are NLI estimation and spatial power profile (SPP) computation for each channel and Raman pump. NLI estimation speed has been slashed by fast closed-form GN/EGN models (CFMs). We have presented here a method that tackles the SPP estimation by greatly speeding it up, without losing accuracy. We have shown its potential by performing a C+L+S and a C+L+S+E 200-channel system optimization, with 3 Raman pumps. Overall, the system GSNR assessment was sped up by a factor of about 100. Thanks to such speed-up, deep optimization was possible, achieving a system configurations with high overall system performance. The C+L+S+E in particular achieved a 4.5x throughput increase with respect to a super-C system, while also providing good GSNR flatness. 
\begin{figure}
    \centering
    \includegraphics[width=1.0\linewidth]{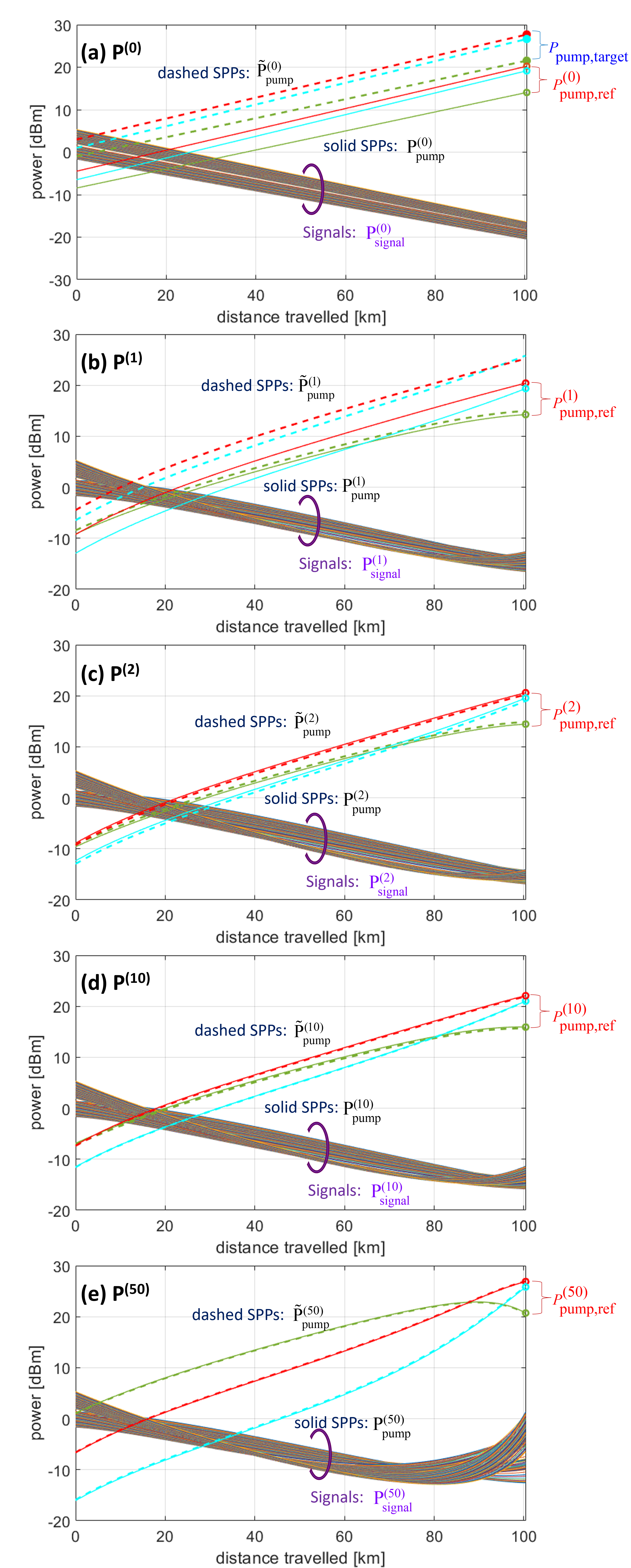}
    \caption{SPPs evolution in CLS system with throuhgput maximization: (a) SPPs initialization $\mathbf{P}^{(0)}$ without considering ISRS. The three dots represent the nominal pump power; (b) SPPs $\mathbf{P}^{(1)}$ after the 1st iteration;(c) SPPs $\mathbf{P}^{(2)}$ after the 2nd iteration; (d) SPPs $\mathbf{P}^{(10)}$ after the 10th iteration; (e) SPPs $\mathbf{P}^{(50)}$ after the 50th iteration. After 75 iterations, SPPs converge to the solution in Fig.~\ref{fig:3bands_SPP}(a).}
    \label{fig:SPPs_calculation} 
\end{figure}
\begin{appendix}
\subsection{pseudo-code for implementing the algorithm}
\label{sect:pseudo-code}
The implementation and general algorithm description has been described in Sect.~\ref{sec:SPPs_algorithm}. Here we show the algorithm in detail, by means of pseudo-code, containing the initialization and the iterative SPPs calculation.

In this section, the CLS system with throughput maximization in Fig.~\ref{fig:3bands_max_throughput}(a)  is used to demonstrate the implementation of the proposed algorithm. A few SPPs are reported in Fig.~\ref{fig:SPPs_calculation}.

\subsubsection{initialization}
The initial SPPs are crucial, especially for BRPs SPPs. In this case study, the sum of the launch powers of the BRPs at $z\!=\!L_{\rm span}$ is 7.5 dB higher than the sum of the launch powers of the transmission channels at $z\!=\! 0$. 

If no modification is imposed on the BRPs, similar to \cite{2005_OL_Choi}, the initial SPPs are calculated based on the nominal launch powers of all lightwaves and the intrinsic fiber loss, which generates $\big[\mathbf{P}_\text{signal}^{(0)};\mathbf{\tilde P}_\text{pump}^{(0)}\big]$ in Fig.~\ref{fig:SPPs_calculation}(a). When using these SPPs as $\mathbf{P}^{(0)}$, the algorithm would diverge immediately due to the significant power transfer from BRPs to signals at the fiber end. 

We therefore proposed to scale down BRPs first and then scale it up slowly so that the interaction between BRPs and signals is mild in each step. The law is to make the two sums coincide initially. In this case study, the scaling factor $t_{s,\text{dB}}$ is 7.5 dB, pushing $\mathbf{\tilde P}_\text{pump}^{(0)}$ down to $\mathbf{P}_\text{pump}^{(0)}$ in Fig.~\ref{fig:SPPs_calculation}(a), yielding a good initial SPPs  $\mathbf{P}^{(0)} = \big[\mathbf{P}_\text{signal}^{(0)}; \mathbf{P}_\text{pump}^{(0)} \big]$.

We also define the number of iterations $ L_\text{iter} $ based on the law that 10 dB is recovered in 100 iterations, where a linearly decreasing step size $\delta_\text{dB}^{(k)}$ is used with $\sum_{k=1}^{k=L_\text{iter}}\delta_\text{dB}^{(k)} = t_{s,\text{dB}}$. $\delta_\text{dB}^{(k)}$ gets smaller as the BRPs become closer to their launch powers, which can help SPPs to converge steadily. In this case study, $ L_\text{iter} $ is 75. $\delta_\text{dB}^{(k)}$ is set as 0.2 dB at the first iteration, decreasing linearly to 0 dB at the last iteration.

Notably, $\delta_\text{dB}^{(k)}$ simultaneously defined BRPs powers reference ${P}_\text{pump,ref}^{(k)}$ for each iteration, as the circles shown in Fig.~\ref{fig:SPPs_calculation}. As mentioned in Sect.~\ref{sec:SPPs_algorithm}, the forward-nature of the algorithm does not impose any constraint on the SPP values at $z\!=\!L_{\rm span}$, we need to rescale BRPs SPPs to gradually approach the launch powers. The $k$-th BRPs reference is,
\begin{equation*}
    {P}_\text{pump,ref}^{(k)} = {P}_\text{pump,ref}^{(k-1)} + \delta_\text{dB}^{(k)}
\end{equation*}  
with the starting values of ${P}_\text{pump,ref}^{(0)}$.

\subsubsection{SPPs calculation}
After the initialization, loading $\mathbf{P}^{(0)}$ into Eq.~(\ref{eq:matrix_power_profile}) generates $\mathbf{\tilde P}^{(1)} = \big[\mathbf{P}_\text{signal}^{(1)}; \mathbf{\tilde P}_\text{pump}^{(1)} \big]$ in the forward direction. As ISRS starts to count, $\mathbf{\tilde P}^{(1)}$ deviate significantly from $\mathbf{P}^{(0)}$. The signals gain some power from BRPs and start to bend up slightly. The BRPs powers at $z\!=\!L_{\rm span}$ are completely off from ${P}_\text{pump,ref}^{(1)}$ as it is impossible to impose any constraint at $z\!=\!L_{\rm span}$. We then rescale $\mathbf{\tilde P}_\text{pump}^{(1)}$ to match with ${P}_\text{pump,ref}^{(1)}$ at $z\!=\!L_{\rm span}$ and obtain $\mathbf{P}^{(1)}$. This completes the first iteration of the algorithm. All SPPs are shown in Fig.~\ref{fig:SPPs_calculation}(b). 

Repeating the same procedure can generate $\mathbf{P}^{(2)}$ (Fig.~\ref{fig:SPPs_calculation}(c)), and so on. Fig.~\ref{fig:SPPs_calculation} displays also $\mathbf{P}^{(10)}$ (Fig.~\ref{fig:SPPs_calculation}(d)) and $\mathbf{P}^{(50)}$ (Fig.~\ref{fig:SPPs_calculation}(e)), demonstrating a stable iteration towards the final solution. After 75 iterations, all BRPs get back to the launch powers and the algorithm stops. Generally, the last ten $\delta_\text{dB}^{(k)}$ are quite small for fine-tuning the solution.

If the algorithm can not deal with the problem, it would diverge within a few iterations, and all SPPs values would accumulate quickly to NaN. We then move to the conventional method. However, this did not happen in all the optimizations in this paper. 

\begin{algorithm}
\FloatBarrier
\caption{ SPPs calculator }
\begin{algorithmic}
    \State \textbf{INITIALIZATION:} 
    \State (1) $\mathbf{P}^{(0)} = \big[\mathbf{P}_\text{signal}^{(0)}; \mathbf{P}_\text{pump}^{(0)} \big]$ 
    \vspace{0.3em}
    \State \hspace{1.5em} \text{SPPs are calculated based only on fiber loss}
    \State \hspace{1.5em} \text{BRPs SPPs are scaled down by a factor of $t_{s,\text{dB}}$}
    \State \hspace{1.5em} \text{to coincide the total power with that of signals}
    \vspace{0.5em}
    \State (2) $ L_\text{iter} \text{: number of iterations} $
    \vspace{0.5em}
    \State (3) $\delta_\text{dB}^{(k)} \text{: the $k$-th step for generating  ${P}_\text{pump,ref}^{(k)}$ while  }$
    \State \hspace{1.5em} \text{linearly decreasing along the iterations}
    \vspace{1em}
    \State \textbf{START SPPs calculation:}
    \vspace{0.3em} 
    \While {$ k <= L_\text{iter} $}
        \vspace{0.3em} 
        \State \text{1. run forward Eq.~(\ref{eq:matrix_power_profile}):} 
        \State \hspace{0.2em} $\mathbf{\tilde P}^{(k)}=\big[\mathbf{P}_\text{signal}^{(k)}; \mathbf{\tilde P}_\text{pump}^{(k)} \big]$
        \vspace{0.8em} 
        \If{\text{NaN appears in SPPs}}
        \State \text{Diverge, and move to the conventional method}
        \EndIf
        \vspace{0.5em} 
        \State \text{2. prepare the $k$-th BRPs reference $\mathbf{P}_\text{pump,ref}^{(k)}$:}
        \State \hspace{0.2em} ${P}_\text{pump,ref}^{(k)} = {P}_\text{pump,ref}^{(k-1)} + \delta_\text{dB}^{(k)}$
        \vspace{0.5em} 
        \State \text{3. rescale $\mathbf{\tilde P}_\text{pump}^{(k)}$ w.r.t. $\mathbf{P}_\text{pump,ref}^{(k)}$:} 
        \State \hspace{0.2em} $ \mathbf{P}_\text{pump}^{(k)} = \mathbf{\tilde P}_\text{pump}^{(k)} / \mathbf{\tilde P}_\text{pump,out}^{(k)} \cdot \mathbf{P}_\text{pump,ref}^{(k)}$
        \vspace{0.5em} 
        \State \text{4. update $\mathbf{P}^{(k)}$:} 
        \State \hspace{0.2em} $\mathbf{ P}^{(k)}=\big[\mathbf{P}_\text{signal}^{(k)}; \mathbf{P}_\text{pump}^{(k)} \big]$
        \vspace{0.5em}
    \EndWhile
\end{algorithmic}
\label{SPPs_code}
\FloatBarrier
\end{algorithm}

\end{appendix}


\end{document}